\documentclass{elsarticle}

\usepackage{graphicx}
\usepackage{epsfig}
\usepackage{fancyhdr}
\usepackage{empheq}
\usepackage{mathbbol}
\usepackage{wasysym}
\usepackage{pstricks}
\usepackage{color}
\usepackage{bbm}

\begin{document}

\title{Multi-GeV Electron Spectrometer}

%
\author[7,8]{R.Faccini
\footnote{rfaccini@slac.staford.edu, Dip. Fisica, "Sapienza" University of Rome,  Ple A. Moro 2, 00185 Rome, Italy, tel +39 0649914798, fax +39 064957697} }
\author[3]{ F.Anelli}
\author[11]{ A. Bacci}
\author[12]{ D. Batani}
\author[3]{ M. Bellaveglia}
\author[12]{ R. Benocci}
\author[5,6]{C.Benedetti}
\author[3]{ L. Cacciotti}
\author[1,2]{C.A.Cecchetti}
\author[3]{A.Clozza}
\author[3]{L. Cultrera}
\author[3]{G.Di~Pirro}
\author[7,8]{ N.Drenska}
\author[3]{ F.Anelli}
\author[3]{M. Ferrario}
\author[3]{ D. Filippetto}
\author[3]{ S.Fioravanti}
\author[3]{ A. Gallo}
\author[1,2]{A.Gamucci}
\author[3]{ G. Gatti}
\author[3]{ A. Ghigo}
\author[1,2]{A.Giulietti}
\author[1,2,3,9]{ D.Giulietti}
\author[1,2,3]{L.A.Gizzi}
\author[1,2]{P. Koester}
\author[1,2]{  L.Labate}
\author[1,3]{T.Levato}
\author[3]{ V.Lollo}
\author[6,10]{ P. Londrillo}
\author[7,8]{S. Martellotti} 
\author[3]{E.Pace}
\author[1,9]{ N.Patack}
\author[11]{ A. Rossi}
\author[7,8]{  F. Tani} 
\author[11]{ L. Serafini}
\author[5,6]{G.Turchetti}
\author[3]{ C.Vaccarezza}
\author[7]{ P.Valente}

\address[7]{ Sez. INFN Roma-1, Roma, Italy}
\address[8]{Dip. Fisica, Univ. La Sapienza, Roma, Italy}
\address[3] {LNF, INFN, Frascati, Italy}
\address[11]{Sez. INFN and Dip. di Fisica, Univ. di Milano, Italy}
\address[12]{Sez. INFN and Dip. di Fisica, Univ. di Milano-Bicocca, Italy}
\address[5]{Dip. di Fisica, Univ. di Bologna, Italy}
\address[6]{Sez. INFN Bologna, Bologna, Italy}
\address[1]{ ILIL-IPCF, CNR, Pisa, Italy}
\address[2] {Sez. INFN, Pisa, Italy}
\address[10]{ Dip. Astronomia, Univ. di Bologna and INAF sezione di Bologna, Italy}
\address[9]{Dip. di Fisica, Univ. di Pisa, Italy}

\begin{abstract}
The advance in laser plasma acceleration techniques pushes the regime of the resulting accelerated particles to higher energies and intensities. In particular the upcoming experiments with the FLAME laser at LNF will enter the GeV regime with almost 1pC of electrons. 
From the current status of understanding of the acceleration mechanism, relatively large angular and energy spreads are expected. There is therefore the need to develop a device capable to measure the energy of electrons over three orders of magnitude (few MeV to few GeV) under still unknown angular divergences.
Within the PlasmonX experiment at LNF a spectrometer is being constructed to perform these measurements. It is made of an electro-magnet and a screen made of scintillating fibers for the measurement of the trajectories of the particles. The large range of operation, the huge number of particles and the need to focus the divergence present unprecedented challenges in the design and construction of such a device.
We will present the design considerations for this spectrometer and the first results from a prototype.

\end{abstract}

\maketitle

\thispagestyle{fancy}




\section{Introduction.}
\label{sec:intro}
The aim of the PLASMONX~\cite{CDR} at the LNF (Laboratori Nazionali di Frascati)  is to develop, with a world-class, high-power laser facility, an innovative, high-gradient acceleration technique based upon super-intense and ultra-short laser pulses and the SPARC experiment~\cite{sparc}, and X/$\gamma$-ray tunable sources using Thomson scattering of optical photons by energetic electrons. 
\begin{figure}[htb]
\begin{center}
\psfig{file=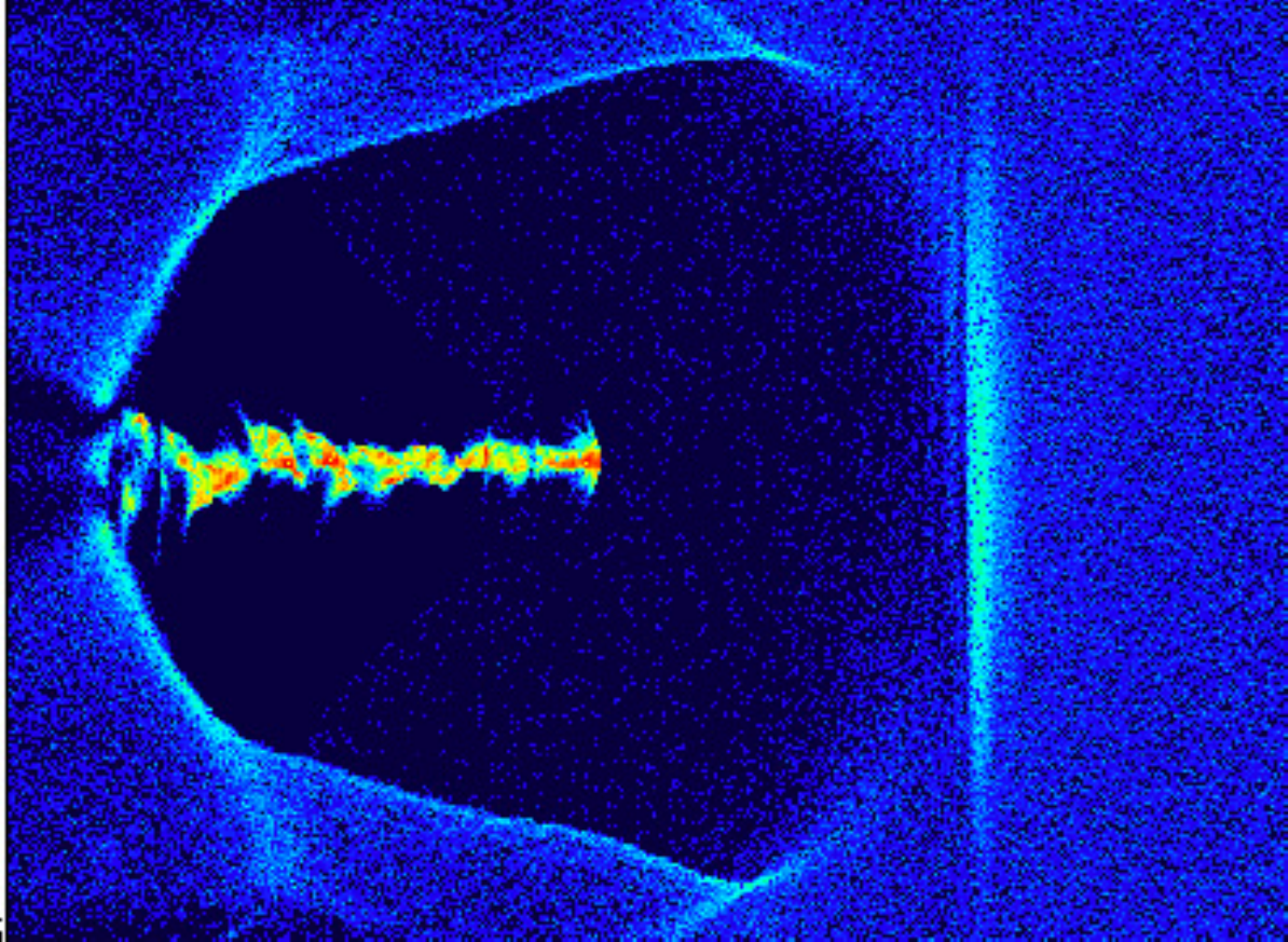,width=5cm}
\psfig{file=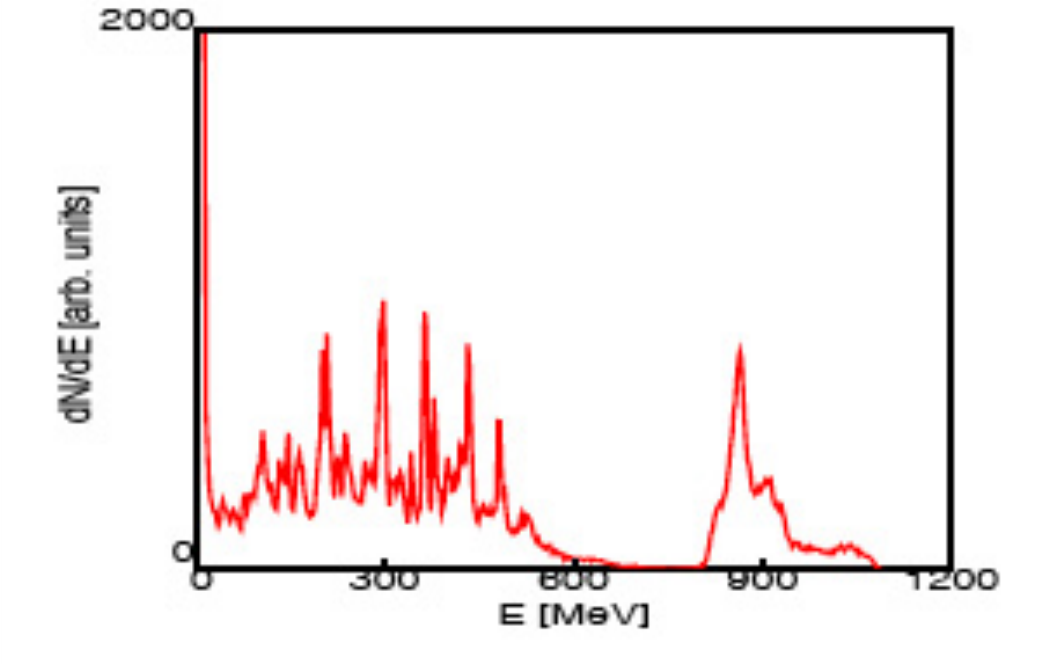,width=5.8cm}
\end{center}
\caption{Results of the 3D PIC simulation: (left) electron density distribution shortly before the end of the acceleration and (right) outcoming energy spectrum.}
\label{fig:PIC}
\end{figure}

The first step of such a program is a test esperiment of self-injection (SITE),  where the laser is focussed on a gas-jet with the goal of producing sub-GeV-class electron bunches from laser-plasma interactions. 
The increase in the energy of the accelerated particles opens the field to high-energy specific diagnostic detectors. In particular the experiment will also be equipped with a magnetic spectrometer. The principle of work is simple: the accelerated charged particles (electrons or ions) will traverse a region where a magnetic field is present and will therefore have a momentum-dependent trajectory. The measurements of the position of such particles on appropriately shaped detectors allows the measurement of the momentum.

The SITE experiment is described in detail elsewhere~\cite{SITE}. To design the spectrometer  it is important to know the characteristics of the electron bunches. To this aim a 3D PIC simulation was performed with plausible laser and gas parameters:  pulse duration = 30 fs, pulse energy = 5 J, peak pulse power = 166 TW, spot size radius $= 9\mu$m, maximum neutral atomic density=1.5 10$^{18} \rm W/cm^2$, and gas depth = 4mm. The resulting electron density in Fig.~\ref{fig:PIC}  towards the end of the acceleration shows that the experiment expects to test the bubble regime. The corresponding energy spectrum  in Fig.~\ref{fig:PIC} shows two components, a  peak with total charge $\sim 700$pC at high energy (around 900 MeV) and a  $\sim 3$nC  low energy tail.

\begin{figure}[htb]
\begin{center}
\psfig{file=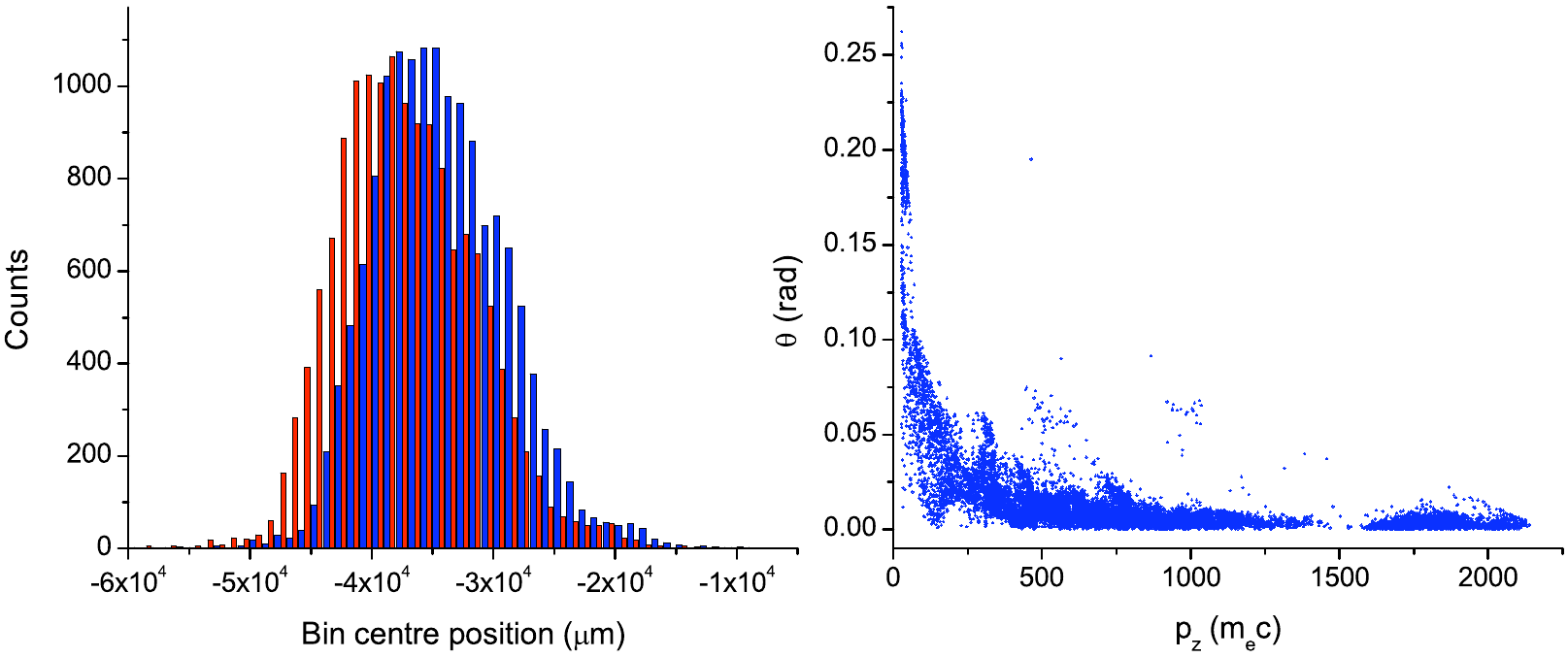,width=10cm}
\end{center}
\caption{(left) Impact of the bunch transport from the exit of the plasma to the spectrometer on the position spectrum and (right) angular divergence  of the beam at the end of the acceleration as  a function of the momentum.}
\label{fig:retar}
\end{figure}

From the high-energy point of view, this detector represents a challenge because the requirements are unprecedented for the field: it must measure the momentum spectrum of tens of millions of particles arriving simultaneously spread over three order of magnitudes in momentum (10 MeV to 10 GeV). 
Since the goal of the experiment is to find the optimal configuration to reduce the low energy tail and shrink the distribution of the high energy portion of the spectrum, the whole spectrum needs to be retained for all bunches. It is therefore not foreseen any form of focussing before the spectrometer (e.g. quadrupoles) since this would select only portions of the momentum spectrum.

The transport of the electron bunches from the exit of the plasma to the spectrometer has been simulated with a 3D parallel tracking code for the beam dynamics of charged particles~\cite{retar} in order to test possible collective effects in such high charge bunches. The position of the electrons when impacting on the optimized detector of the spectrometer as described later is shown in Fig.~\ref{fig:retar} with and without considering the electromagnetic interactions among the electrons. The effect is negligible, the resolution being dominated by the angular divergence of the beam at origin.

For the purpose of designing the spectrometer, which is insensitive to the vertical position of the particles and to their direction of flight when impacting the detector,  it is therefore sufficient to consider the motion of the particles on the horizontal plane (i.e. the one perpendicular to the magnetic field). This justifies why in the following the bunch is treated as a beam of independent particles coming from a point-like source located at the end of the gas-jet and with an angular divergence estimated with the simulation. Fig.~\ref{fig:retar} shows that the high energy particles are contained within 2mrad, while the low energy tail can have a significantly larger divergence.

In summary, the requirements for this detector is to be able to reconstruct the full spectrum from 10 MeV to several GeV, considering a beam coming from a point-like source about a meter upstream and with a divergence of approximately 2mrad in the region of larger interest.

The urgency of this project (which started just a year before the first expected laser-plasma acceleration experiments) forced the use of mostly reused equipment, in particular the magnet is a spare from the SPARC experiment. The results shown in this paper are therefore referred to a non-optimal magnetic setup, that can be seen in Fig.~\ref{fig:focii}.

\section{Design of the Spectrometer.}
Several aspects need to be considered when designing the spectrometer: where and how to impact the magnetic field, where to locate the position detectors, which solution to use for the detectors and how to implement the readout electronics. 

\subsection{Geometry of magnet and detectors}

The shape of the magnetic field and the location of the position detectors are dictated by the need to minimize the resolution on the momentum for a fixed resolution on the position coordinate. 
To perform this optimization, the detector response is estimated by means of 
a Runge-Kutta integration of motion that takes into account the geometry and the detailed magnetic field map. For the reasons detailed in Sec.~\ref{sec:intro},  it is sufficient for the purpose of the detector optimization  to sample two trajectories coming from the point-like source and thereby oriented at $\pm 1\sigma$ where $\sigma$ is the expected beam divergence on the horizontal plane (see Fig.~\ref{fig:retar}). 

\begin{figure}[h!]
\begin{center}
\psfig{file=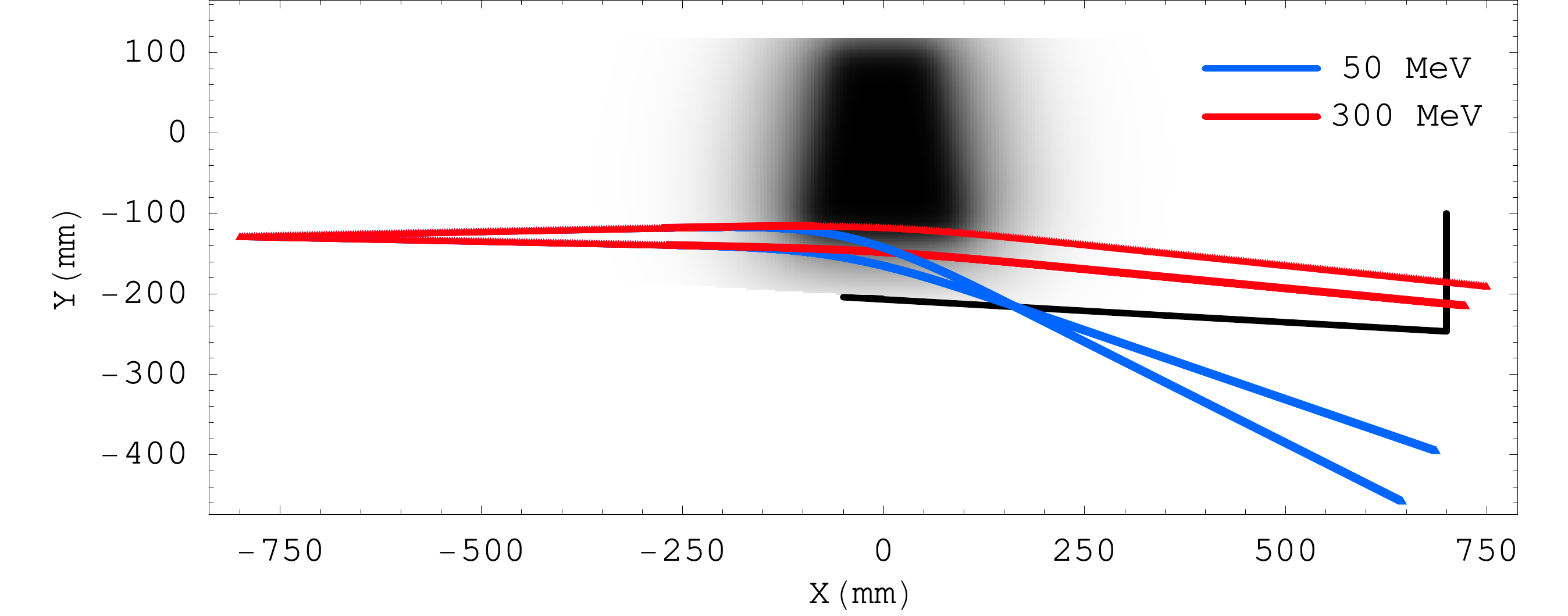,width=12cm}
\end{center}
\caption{Trajectories in the available magnetic setup of particles with fixed momenta and divergence angle $=\pm 2$mrad. The position detectors are located along the black lines.}
\label{fig:focii}
\end{figure}

While the optimal dispersion can be achieved by positioning the detectors as far as possible, this is not necessarily the optimal criterium. The momentum resolution is in fact dominated by the angular dispersion which overlays in a given position trajectories from significantly different momenta. This component of the error can be shrunk if there exist focii of the trajectories, i.e. points where all trajectories of a given momentum converge regardless of the angle at the origin. We found that for the uniform magnetic fields as we are currently investigating, this can be achieved if the magnetic field is not null in regions elongated along the direction of the impacting beam. For the available magnet and hardware configuration, focii were found for momenta up to 150MeV(See Fig.~\ref{fig:focii} ).

With the currently available setup the best configuration is to set the distance of the beam from the magnetic center to 13cm, i.e. to have the trajectories develop mostly in the fringe field, posing operational difficulties that will be discussed later.
Due to these considerations the low momentum detector has been positioned on the locus of focii, while the high momentum detector has been positioned as far as possibile. The final setup is shown in Fig. ~\ref{fig:focii}.

\section{Choice of detector and readout }
Usually in Laser Plasma Acceleration experiments optical devices are used to analyze the electron energy spectrum, in particular positions are measured by LANEX films read by CCD cameras. In this case there are several disadvantages of such a solution: the large size of the magnet makes it difficult to tune the optical setup for such a detector, syncrotron radiation photons are also detected and risk to bias the charged particle spectra, the large energies reached by the accelerated particles risk to saturate the device

\begin{figure}[htb]
\begin{center}
\psfig{file=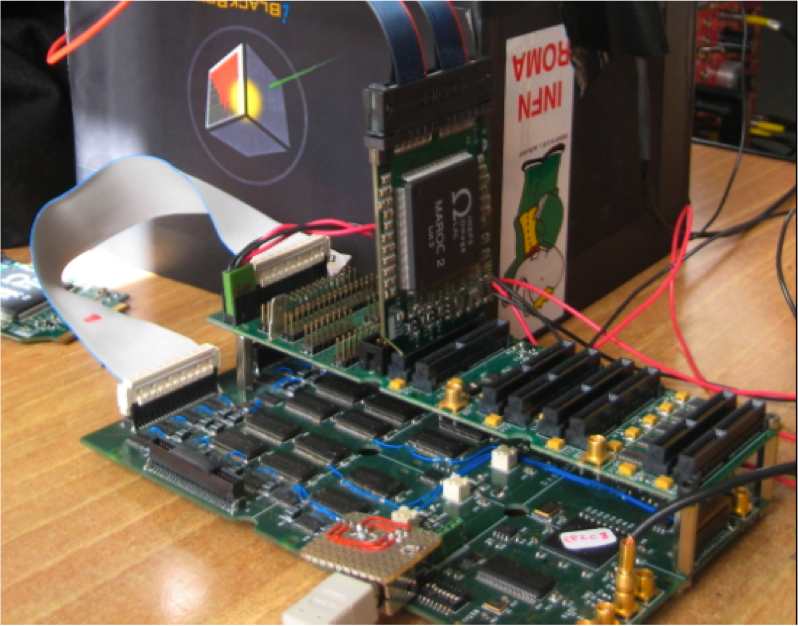,width=6cm}
\psfig{file=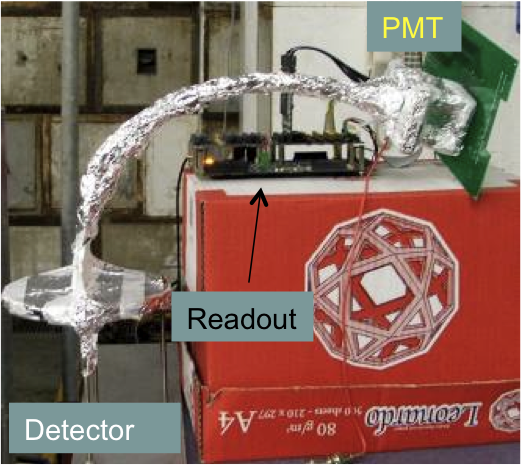,width=5.2cm}
\end{center}
\caption{Left: electronic setup. Right: complete prototype of the position detectors.}
\label{fig:maroc}
\end{figure}

Among the non-optical devices, we had to choose a detector that could operate in vacuum, tolerate a large number of impacting particles and be of limited cost. We therefore chose to use scintillating fibers.  The scintillating fibers Kuraray SCSF-81-SJ with diameter 1.00 $\pm$ 0.05 mm have 50 $\pm$ $\mu m $ thickness of cladding and emission wavelength 437 nm. Fibers are connected to five multi-channel photo-tubes Hamamatsu H7545 (R7600-00) for a total of five PMT and 320 electronic channels. In order to read a larger number of fibers, since the overall resolution is not affected, the fibers coming from the low momentum detector are merged in groups of three.
The choice of the front end card was studied in collaboration with INFN BA, GE and  ISS-Roma1~\cite{maroc}. It is possible to multiplex up to 4096 channels thanks to the technology based on MAROC2 chips shown in Fig.~\ref{fig:maroc}.

\section{Data analysis}
A signal proportional to the number of impacting electrons is recorded on each fiber of the detectors.
This information is used to reconstruct the energy spectrum by means of a bayesian unfolding. If we divide the momentum range into $N_p$ bins, the best estimate of the charge deposited in the $p$-th bin, starting from the charge $Q_i$ deposited in the $i$-th fiber, is 
\begin{eqnarray}
\hat{Q}_p&=&\sum_i Q_i \times P(p|i) \\
P(p|i)&=&P(i|p)\times {\cal{P}}_p/\sum_p P(i|p)\times {\cal{P}}_p,
\end{eqnarray}
where $P(p|i)$ are the estimated probability distribution of the momentum if a given fiber, $i$ is hit,
${\cal{P}}_p$ is the a-priori momentum distribution, set by default to flat, and P(i|p) is the detector response, the probability of a particle of momentum $p$ to hit fiber $i$. Further improvement in the estimate can be achieved by reapplying the equations after replacing the a-priori probabilities with the calculated $P(p|i)$. 
\subsection{Estimated detector performances}
\begin{figure}[htb]
\begin{center}
\psfig{file=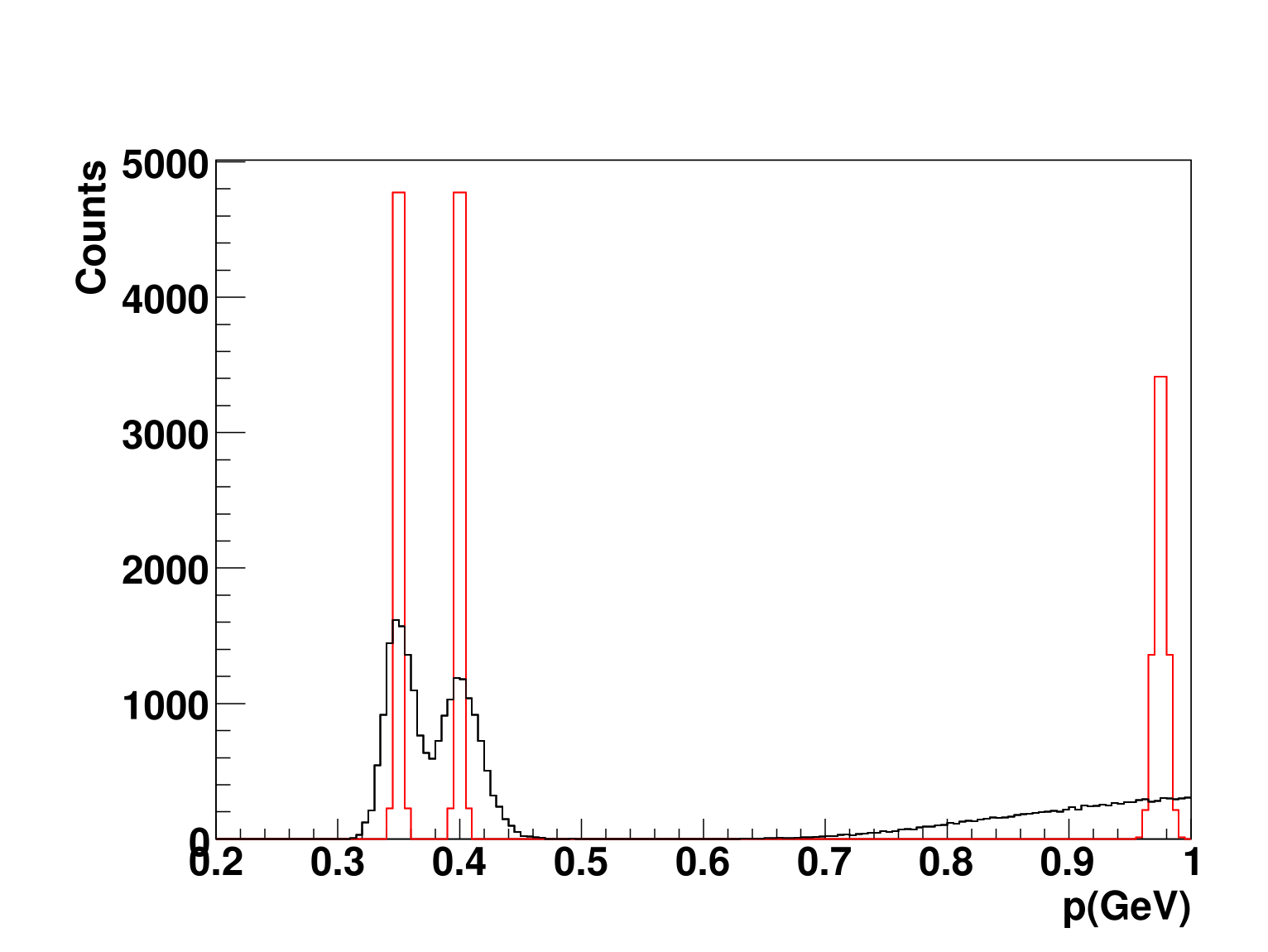,height=3.5cm}
\psfig{file=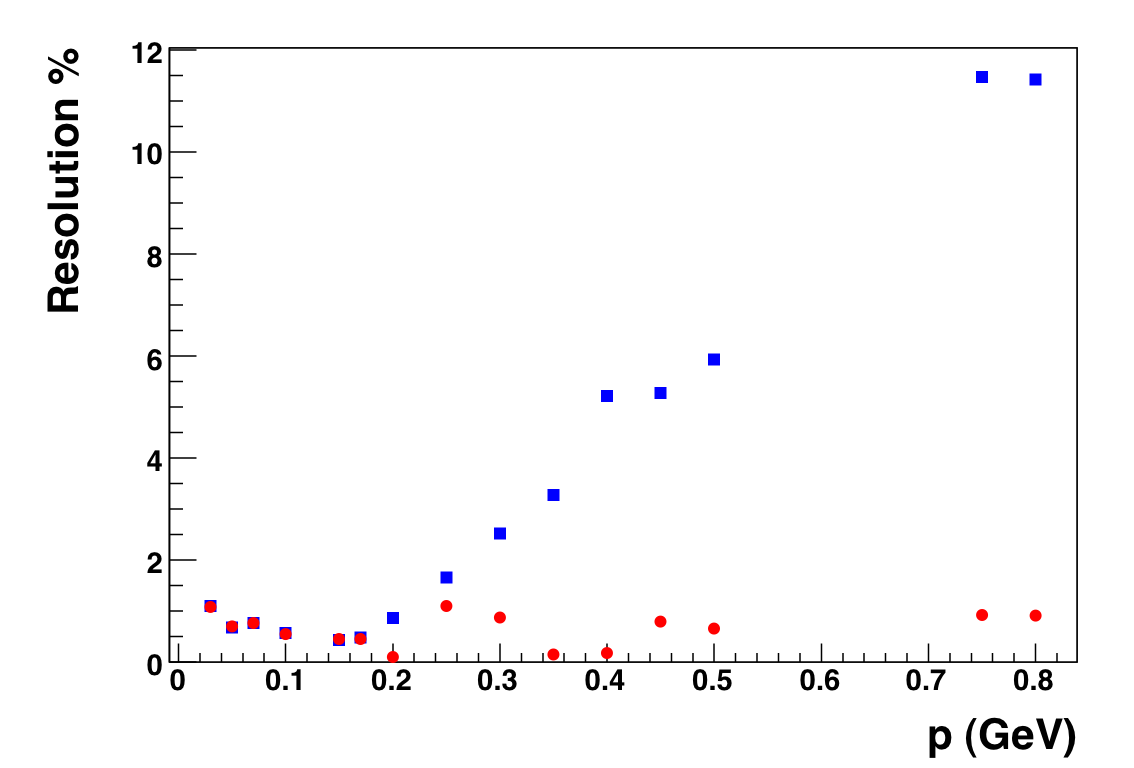,height=3.1cm}
\end{center}
\caption{Left: generated (red) and reconstructed (black) momentum distributions overlayed. 
Right: resolution on momentum (blue squares) and the contribution from the position measurement only (red dots). The difference between the two is an estimate of the 
angular divergerce.}\label{fig:bayes}
\end{figure}
Fig~\ref{fig:bayes} shows the effect of resolution from a fast simulation of almost mono-energetic beams. The worsening of the reconstruction with increasing momentum is clear.

We also estimate the resolution of the detector, separating the 
 intrinsic detector resolution caused by the physical dimension of the fibers (1mm of diameter)
 and the angular divergence. Fig.~\ref{fig:bayes} shows the total resolution as a function of the momentum and the component due to the detector granularity.
 Performances between low and high detector will be different mainly because of the angular effect for the high energies.
The pointing instability of the laser was also studied and our results show that the situation shown above is not significantly different even if these effects are of the order of degree.
We can conclude that with the Prototype it is possible to measure energies up to 200MeV with a very good resolution less then 1\% and the resolution remains less then 5\% up to about 500Mev.

\section{Tests with a prototype}

A prototype with 64 fibers read by one PMT (see Fig.~\ref{fig:maroc}) has been built and tested with electrons at the Frascati Beam Test Facility (BTF~\cite{BTF}) and in laboratory with LEDs.

At the BTF test the electron beams was passing through the magnet that will be part of the detector, so that we could test the existence of the focii and the understanding of the fringe field region. The fact that the trajectories lie in the fringe area  is in fact one the major concerns for this device.

\begin{figure}[htb]
\begin{center}
\psfig{file=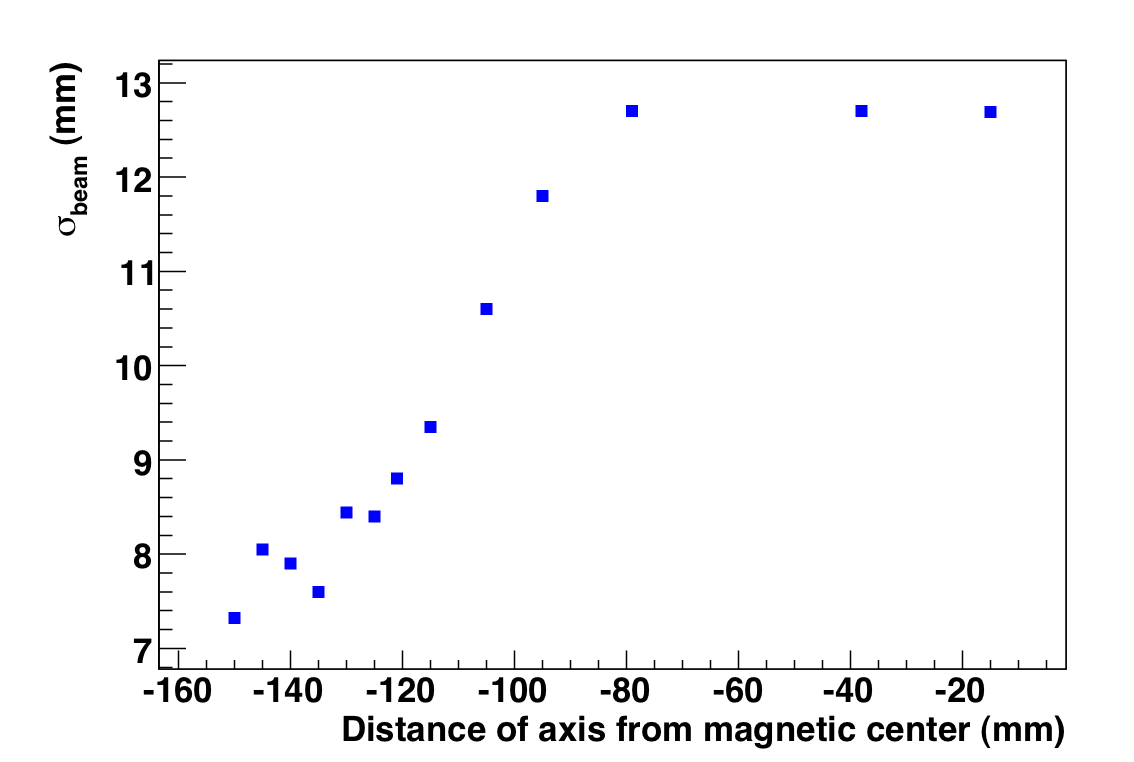,height=3.2cm}
\psfig{file=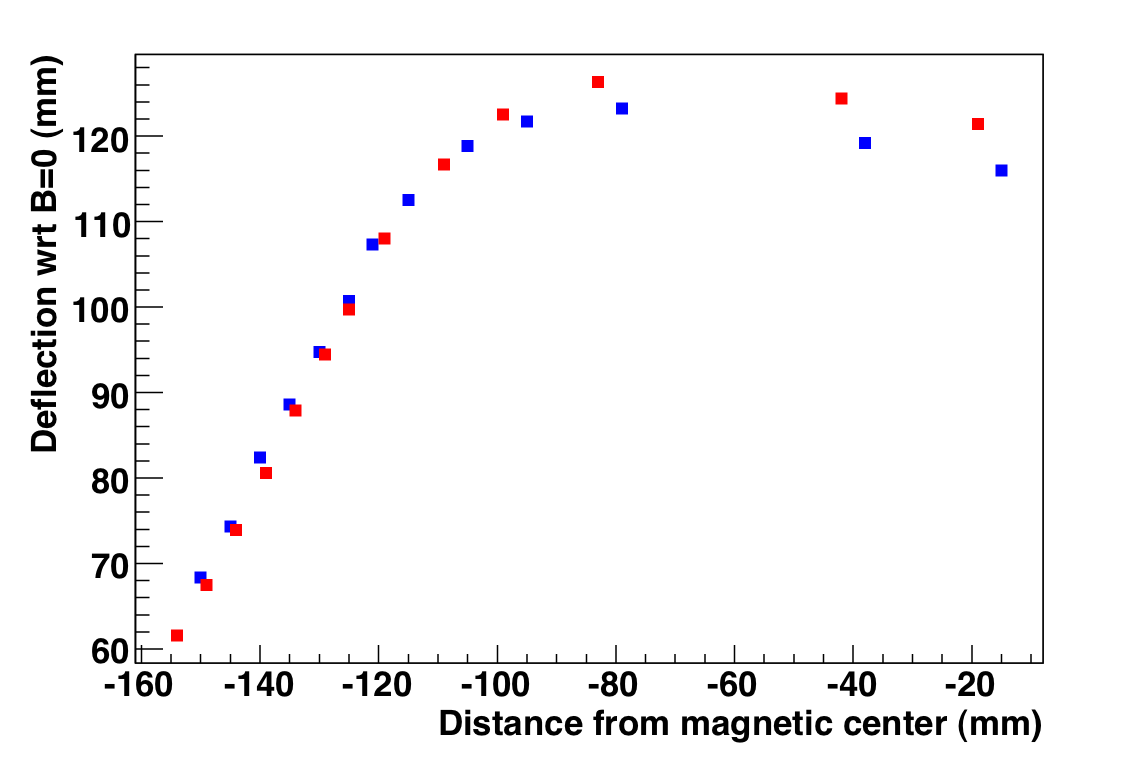,height=3.2cm}
\end{center}
\caption{BTF test beam: (left) measured resolution as a function of the distance 
from the magnetic center and (right) comparison between prediction (red) and measurement (blue) for the beam deflection as a function of the position of the beam wrt the magnetic center. }
\label{fig:wp}
\end{figure}

Fig.~\ref{fig:wp} shows that  moving the beam from the center of the magnet to the fringe, the resolution, due to multiple scattering in air,  improves and that the deviation of the beam as a function of the beam position is well reproduced by the numerical calculation.

\section*{Acknowledgments}
We want to thank Evaristo Cisbani  for his collaboration on the development of the electronics.

\end{document}